\documentclass[journal=./achemso/jpc,manuscript=article,layout=onecolumn]{achemso}
\usepackage{geometry}
\usepackage{amsmath}
\usepackage{amsfonts,amstext,amsgen,amsbsy,amsopn,amssymb}
\usepackage[braket,qm]{qcircuit}
\usepackage{graphicx}
\usepackage{dcolumn}
\usepackage{bm}
\usepackage{physics}
\usepackage{makecell}
\usepackage{booktabs}
\usepackage{caption}
\usepackage{booktabs} 

\usepackage{subcaption}
\usepackage{multirow}
\usepackage[dvipsnames, table]{xcolor}
	\definecolor{diagramColor}{RGB}{0,0,120}
\usepackage{tikz}
	\usetikzlibrary{trees}
	\tikzset{
 		treenode/.style = {shape=rectangle, draw, diagramColor, ultra thick, align=center, 
                     top color=white, bottom color=white},
		root/.style     = {treenode, font=\normalsize},
  		env/.style      = {treenode, font=\normalsize},
		envtwo/.style      = {treenode, font=\normalsize},
		mode/.style = {edge from parent path={(\tikzparentnode.east) -- (\tikzchildnode.west)}}
	}
\usepackage{geometry}
\usepackage{enumitem}
\usepackage{mathtools}

\usepackage[colorlinks=true,linkcolor=blue,citecolor=blue,urlcolor=black]{hyperref}
\usepackage{array}
\newcolumntype{?}{!{\vrule width 1.5pt}}

\newcommand{\halfcheckmark}[0]{\checkmark\raisebox{0.23em}{\kern-0.68em\large$\times$}}

\SectionNumbersOn

\title{
Simulating Electron Transfer in a Molecular Triad within an Optical Cavity Using NISQ Computers
}

\author{Ningyi Lyu}
\affiliation{Department of Chemistry, Yale University, New Haven, CT 06520, U.S.A.}
\author{Pouya Khazaei}
\affiliation{Department of Chemistry, University of Michigan, Ann Arbor, MI  48109, U.S.A.}

\author{Eitan Geva}
\affiliation{Department of Chemistry, University of Michigan, Ann Arbor, MI  48109, U.S.A.}
\email{eitan@umich.edu}

\author{Victor S. Batista}
\affiliation{Department of Chemistry, Yale University, New Haven, CT 06520, U.S.A.}
\alsoaffiliation{Yale Quantum Institute, Yale University, New Haven, CT 06511, U.S.A.} 
\email{victor.batista@yale.edu}

\date{\today}
\makeatother
\begin{document}

\section{Abstract}
We present a quantum algorithm based on the Tensor-Train Thermo-Field Dynamics (TT-TFD) method to simulate the open quantum system dynamics of intramolecular charge transfer modulated by an optical cavity on noisy intermediate-scale quantum (NISQ) computers. We apply our methodology to a model that describes the $\pi\pi^*$ to CT1 intermolecular charge transfer within the carotenoid-porphyrin-C60 molecular triad solvated in tetrahydrofuran (THF) and placed inside an optical cavity. We find how the dynamics is influenced by the cavity resonance frequency and strength of the light-matter interaction, showcasing the NISQ-based simulations to capture these effects. Furthermore, we compare the approximate predictions of Fermi's Golden Rule (FGR) rate theory and Ring-Polymer Molecular Dynamics (RPMD) to numerically exact calculations, showing the capabilitis of quantum computing methods to assess the limitations of approximate methods. 

\section{Introduction}
Exploring the potential of light-matter interactions to influence chemical reactions has captivated the attention of physical chemists for many years.\cite{tannor88,kosloff89,gordon97,assion98,rabitz00,rice00,levis01,pearson01,rice02,shapiro02,mcrobbie16} Recent experimental advancements have showcased the possibility of utilizing these interactions to control various chemical and physical phenomena, including energy and charge transfer, photochemical processes, photocatalysis.~\cite{andrew00,schwartz11,hutchison12,hutchison13,torma14,flick15,feist15,schachenmayer15,shalabney15,orgiu15,long15,ebbesen16,thomas16,zhong16,herrera16,casey16,sanvitto16,kowalewski16,kowalewski16a,flick17,zhong17,martinezmartinez17,fregoni18,saezblazquez18,flick19,galego19,lather19,schafer19,hoffmann19,hoffmann19a,lacombe19,mandal19,nitzan19,hoffmann20,flick20,gu20,mandal20,chowdhury21,saller21,saller22,basov21} The proposed control mechanisms involve both strong and ultra-strong coupling of electromagnetic microcavity optical modes with the electronic and vibrational degrees of freedom (DOF) of molecules within these cavities. These experimental advances call for the development of
accurate computational methods for simulating
the dynamics of molecular matter inside cavities to reveal fundamental mechanisms and predict various experimental outcomes such as reaction yields and transfer rates. So far, most dynamical simulations have utilized mixed quantum classical (MQC) methods 
which might not describe nuclear quantum effects properly since they
treat part of the system classically.~\cite{chowdhury21} On the other hand, numerically exact methods are now developed to treat vibronic systems with a few tens to a few hundreds nuclear degrees of freedom (DoF),\cite{yan2021} while rapidly advancing quantum computers are expected to push that limit further. In this work, we implement the Tensor-Train Thermo-Field Dynamics (TT-TFD) method\cite{Borrelli2021,Lyu2023,WangJCTC2023} for numerically exact quantum dynamics simulations of molecular systems in optical cavities. Utilizing the Sz-Nagy dilation scheme,~\cite{nagy1970harmonic,hu2020quantum} we demonstrate the TT-TFD method as applied to quantum computing simulations on currently available IBM Noisy Intermediate-Scale Quantum (NISQ) computers.



\begin{figure}[H]
\includegraphics[scale=0.8]{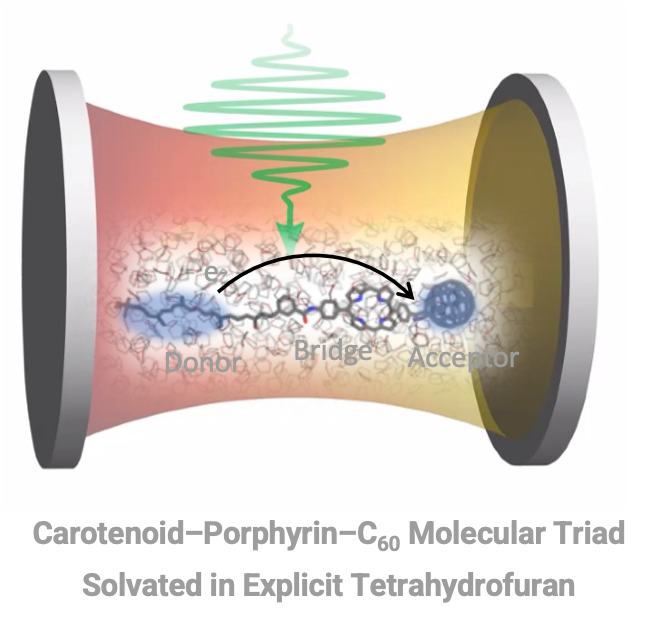}
\caption{Illustration of the THF-solvated molecular triad in cavity.}
\label{fig:triad_cav_TFD1}
\end{figure}

Our simulations investigate electron transfer in the carotenoid-porphyrin-C60 molecular triad dissolved in tetrahydrofuran (THF) inside an optical cavity, as illustrated in Fig~\ref{fig:triad_cav_TFD1}. 
We investigate the intramolecular electron transfer dynamics in the triad under various cavity setups. We find that the dynamics can be controlled significantly by changing the resonance frequency of the cavity, or the strength of the light-matter interaction. Our numerically exact results are compared to calculations based on approximate methodologies, including the Fermi Golden Rule (FGR),~\cite{Saller2023} and Ring-Polymer Molecular Dynamics (RPMD) simulations,~\citenum{chowdhury21} providing valuable insights on the capabilities and limitations of approximate methods.
Therefore, it is natural to expect that our quantum computing methodology should also be useful for other studies of molecular systems in polaritonic cavities. 

\section{Methodology}

\subsection{Hamiltonian for cavity-modified electron transfer}
We consider the following light-matter Hamiltonian,
\begin{equation}\label{eq:Ham}
\hat{H}=-\epsilon |A\rangle\langle A|+\Delta (|D\rangle\langle A|+|A\rangle\langle D|)+\hat{H}_{B}+\hat{H}_F,
\end{equation}
where $|D\rangle$ and  $|A\rangle$ denote the $\pi\pi^*$ and CT1 diabatic states of the solvated triad, respectively, with $\epsilon=532~\text{meV}$ defining the energy gap between the two states and $\Delta=24~\text{meV}$ the electronic coupling.\cite{sun18}

The bath Hamiltonian $H_B$, corresponding to the mass-weighted bath coordinates $R_j$, is described by a set of harmonic oscillators coupled to the two electronic states, as follows:
\begin{equation}
\hat{H}_{B}=
\sum_{j=1}^{50}\bigg(\frac{\hat{P}_j^2}{2}+\frac{\omega_j^2}{2}\hat{R}_j^2\bigg)|D\rangle\langle D|
+
\sum_{j=1}^{50}\bigg(\frac{\hat{P}_j^2}{2}+\frac{\omega_j^2}{2}(\hat{R}_j-\frac{c_j}{\omega_j^2})^2\bigg)|A\rangle\langle A|,
\nonumber
\end{equation}
where the bath frequencies $\{\omega_j\}$ and coupling constants $\{c_j\}$ are obtained by discretizing the following Ohmic spectral density:
\begin{equation}
J(\omega)=\frac{\pi}{2}\sum_{j=1}^{50}\frac{c_j^2}{m_j\omega_j}\delta(\omega-\omega_j)=\eta\omega e^{-\omega/\omega_c},
\end{equation}
where $\omega_c=6.57~\text{meV}$,
and $\eta=1.066~\times 10^6~\text{au}$ are chosen so that the discretized vibrational frequencies roughly cover the range of normal mode frequencies for this system as calculated by molecular dynamics simulations.~\cite{tong20}
The couplings $\{c_j\}$ are then scaled with a constant $a$, such that the values of the scaled $\tilde{c}_j=c_ja$ correspond to the calculated molecular reorganization energy of the bent triad molecule $\lambda=515~\text{meV}$:
\begin{equation}
\frac{1}{2}\sum_{j=1}^{50}\omega_j^2\tilde{c}_j^2=\lambda.
\end{equation}

The coupling of the optical cavity mode with the molecule is described, as follows:
\begin{equation}
\hat{H}_F=\hbar\omega_p\hat{a}_p^\dagger\hat{a}_p+\hbar g_p(|D\rangle\langle A|+|A\rangle\langle D|)(\hat{a}_p^\dagger+\hat{a}_p),
\end{equation}
where $\hat{a}_p$ and $\hat{a}_p^\dagger$ are the photonic annihilation and creation operators, respectively. The values of the light-matter coupling strength ($g_p$) and the cavity frequency ($\omega_p$) determine the effect of the cavity and will be varied to evaluate their effect on the intramolecular electron transfer dynamics. 
\subsection{Tensor-Train Thermo-Field Dynamics Simulations}
The evolution of the quantum state $\rho(t)$ is described by the quantum Liouville equation,
\begin{equation}\label{eq:Liou}
\frac{d}{dt}{\rho}(t)=-\frac{i}{\hbar}[\hat{H},{\rho}(t)],
\end{equation}
which is the equation of motion of the time-evolving density matrix ${\rho}(t)$. The TT-TFD method solves this equation in vectorized form, as follows:
\begin{equation}\label{eq:TFDEoM}
\frac{d}{dt}{\vert \psi(\beta,t)\rangle}=-\frac{i}{\hbar} \bar{H} \vert \psi(\beta,t)\rangle,
\end{equation}
with $\bar{H} = \hat{H} \otimes \tilde{I}$. Here, $\vert \psi(\beta,t)\rangle$ is the thermal wavefunction that defines the density matrix, as follows:
\begin{equation}\label{eq:vvLiou}
\rho(t) =  \text{Tr}_f\{\vert\psi(\beta,t)\rangle \langle \psi(\beta,t)\vert \},\end{equation}
with $\rho(0)$ obtained from Eq.~(\ref{eq:vvLiou}) with the initial thermal distribution: \begin{equation}\vert \psi(0,\beta)\rangle= Z^{-1/2} \sum_n e^{-\beta E_n/2} \vert n, \tilde{n} \rangle. 
\label{eq:0beta}
\end{equation}
Such a transformation effectively recasts Eq.~(\ref{eq:Liou}) into Eq.~(\ref{eq:TFDEoM}) which is the time-dependent Schr\"odinger equation (TDSE)
for the thermal wavefunction. 
Note that according to Eq.~(\ref{eq:0beta}), the thermal wavefunction is expanded in a double Hilbert space $\mathcal{H} \otimes \tilde{\mathcal{H}}$, including states
$|\tilde{n}\rangle$ that are identical copies of states $|n\rangle$ in a Hilbert space $\tilde{\mathcal{H}}$ that is an exact copy of the physical Hilbert space $\mathcal{H}$ of states $|n\rangle$.

The preparation of the initial state $|\psi(\beta,0)\rangle$ as in Eq.~(\ref{eq:0beta}) requires imaginary time propagation, which could be computationally burdensome. However, when the bath Hamiltonian is harmonic, we can define the Hamiltonian, as follows:
\begin{align}
\bar{H} = \hat H \otimes \tilde{I} - \hat I \otimes \sum_j \omega_j \tilde{a}_j^\dagger \tilde{a}_j.
\end{align}
This task is circumvented by utilizing the thermal Bogoliubov transformation,
\begin{align}
|0(\beta)\rangle = e^{-i\hat{G}}|0,\tilde{0}\rangle,
\label{eq:0G}
\end{align}
where
\begin{equation}
\hat{G}=-i\sum_j \theta_j\left(\hat{a}_j\tilde{a}_j-\hat{a}^\dagger_j\tilde{a}^\dagger_j \right),
\label{eq:G}
\end{equation}
with $\theta_j=\text{arctanh}(e^{-\beta\omega_j/2})$ and ${\hat{a}_j,\hat{a}_j^\dagger}$  (${\tilde{a}_j,\tilde{a}_j^\dagger}$) the physical (fictional) bosonic creation and annihilation operators for the $j$-th degree of freedom.

Substituting Eqs.~(\ref{eq:0G}) and~(\ref{eq:G}) into Eq.~(\ref{eq:TFDEoM}), we obtain:
\begin{equation}\label{eq:thetaEoM}
\frac{\mathrm d|\psi_\theta(\beta,t)\rangle}{\mathrm dt}=-\frac{i}{\hbar}\bar{H}_\theta|\psi_\theta(\beta,t)\rangle,
\end{equation}
where:
\begin{equation}
\begin{split}
&\bar{H}_\theta=e^{i\hat{G}}\bar{H}e^{-i\hat{G}},\\
&|\psi_\theta(\beta,0)\rangle=e^{i\hat{G}}|\psi(\beta,0) \rangle = |0,\tilde{0}\rangle.
\end{split}
\end{equation}
For a vibronic system that involves many nuclear degrees of freedom, the thermal wavepacket is a high-dimensional tensor which full representation faces the curse of dimensionality. The tensor train (TT) decomposition\cite{Oseledets2010,Oseledets2011,Lyu2022} is a numerically exact data compression strategy that represents moderately entangled wavefunction with a train of 2- or 3-dimensional tensors, such that they can be stored and computed efficiently. TT-TFD expresses $|0,\tilde{0}\rangle$ as a rank-1 tensor train, and propagates Eq.~(\ref{eq:thetaEoM}) to obtain $|\psi_\theta(\beta,t)\rangle$. The propagation can be carried out with the TT-KSL integrator.~\cite{WangJCTC2023,Lyu2023} The expectation value of physical observables $\hat A$ are obtained, as follows:
\begin{align}
\langle A(t) \rangle = \langle \psi_\theta(\beta,t) | \bar{A}_\theta |\psi_\theta(\beta,t)\rangle,
\end{align}
where $\bar{A}_\theta= e^{i\hat{G}}\bar{A}e^{-i\hat{G}}$ and $\bar{A} = \hat A \otimes \tilde I$.
\subsection{Quantum computation for population dynamics}
Numerically exact benchmark calculations based on TT-TFD simulations of the model carotenoid-porphyrin-C60 molecular triad solvated in tetrahydrofuran (THF) in the optical cavity are performed for the evolution of populations of the electronic states. In what follows we outline the quantum computing procedure for obtaining cavity-modified population dynamics based on the TT-TFD results. 

Our propagation scheme is based on the so-called population-only Liouville space superoperator $\mathcal{P}^{pop}(t)$ that satisfies the following equation:
\begin{equation}\label{vmatvec}
\hat{\sigma}^{\text{pop}}(t)=\mathcal{P}^{\text{pop}}(t)\hat{\sigma}^{\text{pop}}(0),
\end{equation}
where $\hat{\sigma}(t)=\text{Tr}_n[\hat{\rho}(t)]$ is the reduced density matrix for the electronic DOFs, with $\hat{\rho}(t)$ the density matrix for the full vibronic system. Here, $\hat{\sigma}^{\text{pop}}(t)=(\sigma_{00}(t),\sigma_{11}(t))^T$ includes only the diagonal elements of $\hat{\sigma}(t)$, necessary to describe the electronic population dynamics. The preparation of the super-operator $\mathcal{P}^{pop}(t)$ according to the TT-TFD generated population dynamics is described in Appendix E of Ref.~\citenum{lyu2023tensor}. 

To perform quantum computing simulations based on Eq.~\eqref{vmatvec}, we first transform $\mathcal{P}^{\text{pop}}(t)$ into a unitary matrix using the Sz.-Nagy dilation theorem\cite{nagy1970harmonic}, as follows:~\cite{hu2020quantum,levy2014dilation} 
\begin{equation}
\mathcal{U}_{\mathcal{P}^{\text{pop}}}(t)=\begin{pmatrix}
\mathcal{P}^{\text{pop}}(t)&\sqrt{I-\mathcal{P}^{{\text{pop}}}(t)\mathcal{P}^{\text{pop}^\dagger}(t)}\\
\sqrt{I-\mathcal{P}^{\text{pop}{^\dagger}}(t)\mathcal{P}^{\text{pop}}(t)}&-\mathcal{P}^{\text{pop}^\dagger}(t)\\
\end{pmatrix}.
\end{equation}
The vectorized $v_{\sigma(0)}$ is dilated by appending ancillary zero elements, as follows:
\begin{equation}
\hat{\sigma}^{\text{pop}}(0)=(\sigma_{00}(0),\sigma_{11}(0))^T\rightarrow \tilde{\sigma}^{\text{pop}}(0)=(\sigma_{00}(0),\sigma_{11}(0),0,0)^T.
\end{equation}
The dilated time-updated population-only density matrix is obtained, as follows:
\begin{equation}\label{diamatvec}
\tilde{\sigma}^{\text{pop}}(t)=\mathcal{U}_{\mathcal{P}^{\text{pop}}}(t)\tilde{\sigma}^{\text{pop}}(0).
\end{equation} 

The dilation scheme thus provides the unitary matrix $\mathcal{U}_{\mathcal{P}^{\text{pop}}}(t)$ governing the time-evolution of $\tilde{\sigma}^{\text{pop}}(t)$, the first two digits of which agree with those of $\hat{\sigma}^{\text{pop}}(t)$. Therefore, Eqs.~\eqref{diamatvec} and~\eqref{vmatvec} describe the same dynamics, with Eq.~\eqref{diamatvec} allowing for simulations on a NISQ quantum computer. 
For the spin-boson model of interest, $\mathcal{U}_{\mathcal{P}^{\text{pop}}}(t)$ is a $4\times 4$ unitary matrix corresponding to a 2-qubit gate.

\section{Results}

\begin{figure}[H]
\includegraphics[scale=0.7]{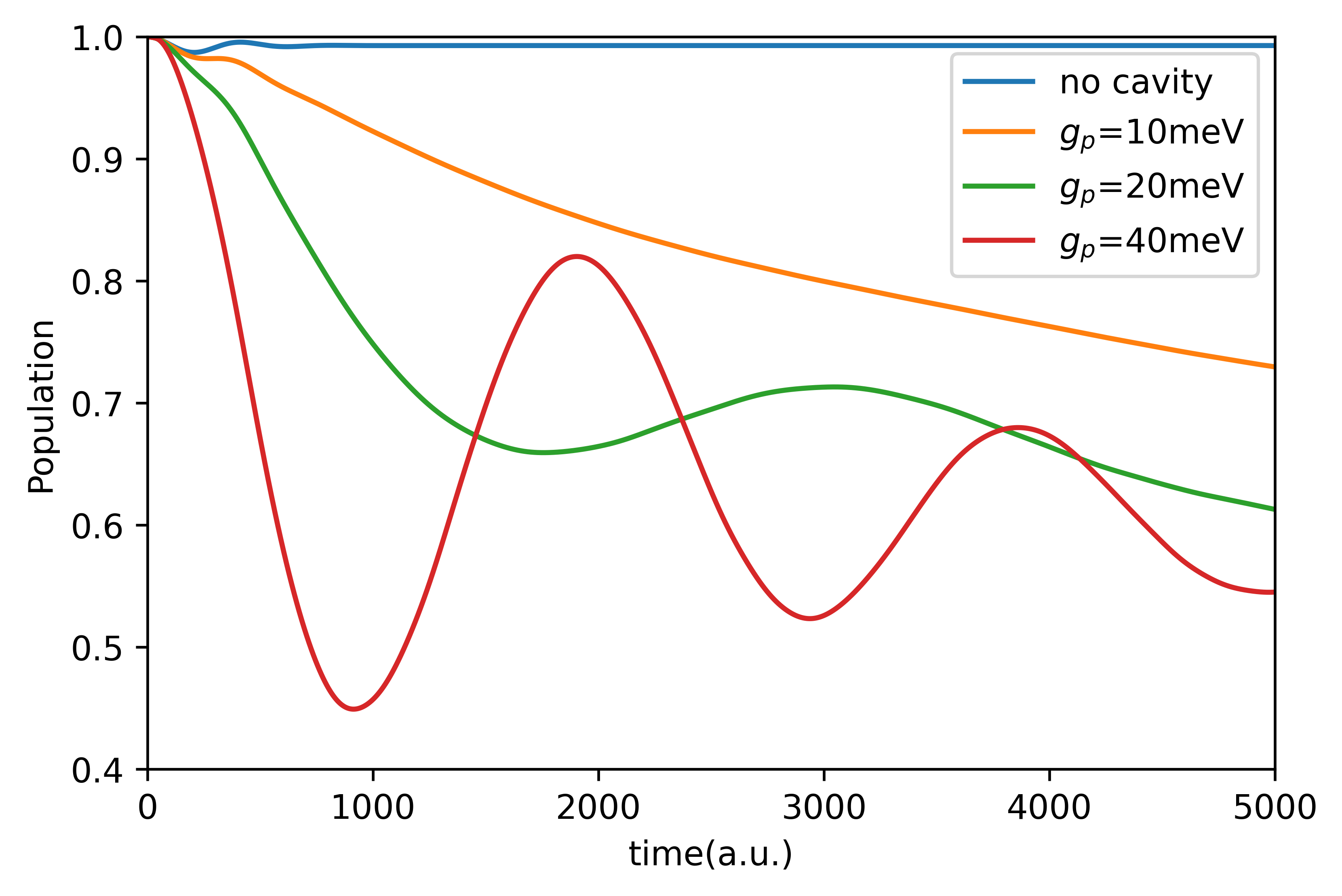}
\caption{Time-dependent population of the ($\pi\pi^*$) state during the early time relaxation after photoexcitation of the carotenoid-porphyrin-C60 molecular triad solvated in tetrahydro-
furan (THF) prepared in a geometry relaxed bent configuration in an optical cavity. The cavity frequency is set as $\omega_p=$~400~meV for all calculations, while the light-matter interaction strength $g_p=$10, 20 and 40~meV. The temperature is set at $T=300~K$. The calculations are based TT-TFD simulations with a propagation time-step $\tau=1~a.u.$
}
\label{fig:triad_cav_TFD}
\end{figure}

Figure~\ref{fig:triad_cav_TFD} shows the calculated short-time population dynamics of the model carotenoid-porphyrin-C60 molecular triad solvated in tetrahydrofuran (THF) in the optical cavity when prepared in a geometry relaxed bent configuration. The results show that in the ultrafast timescale, significant population transfer only happens with the aid of coupling to the cavity mode. Further, increasing the strength of the coupling to the cavity mode changes the electron transfer dynamics qualitatively by becoming highly oscillatory, while in the weaker coupling regime the transfer dynamics exhibits a smooth decay as predicted by rate kinetic theory. The observed oscillatory behavior is likely induced by to Rabi oscillations due to the photonic coupling term with a frequency approximately doubling by doubling $g_p$.

\begin{figure}[H]
\includegraphics[scale=0.8]{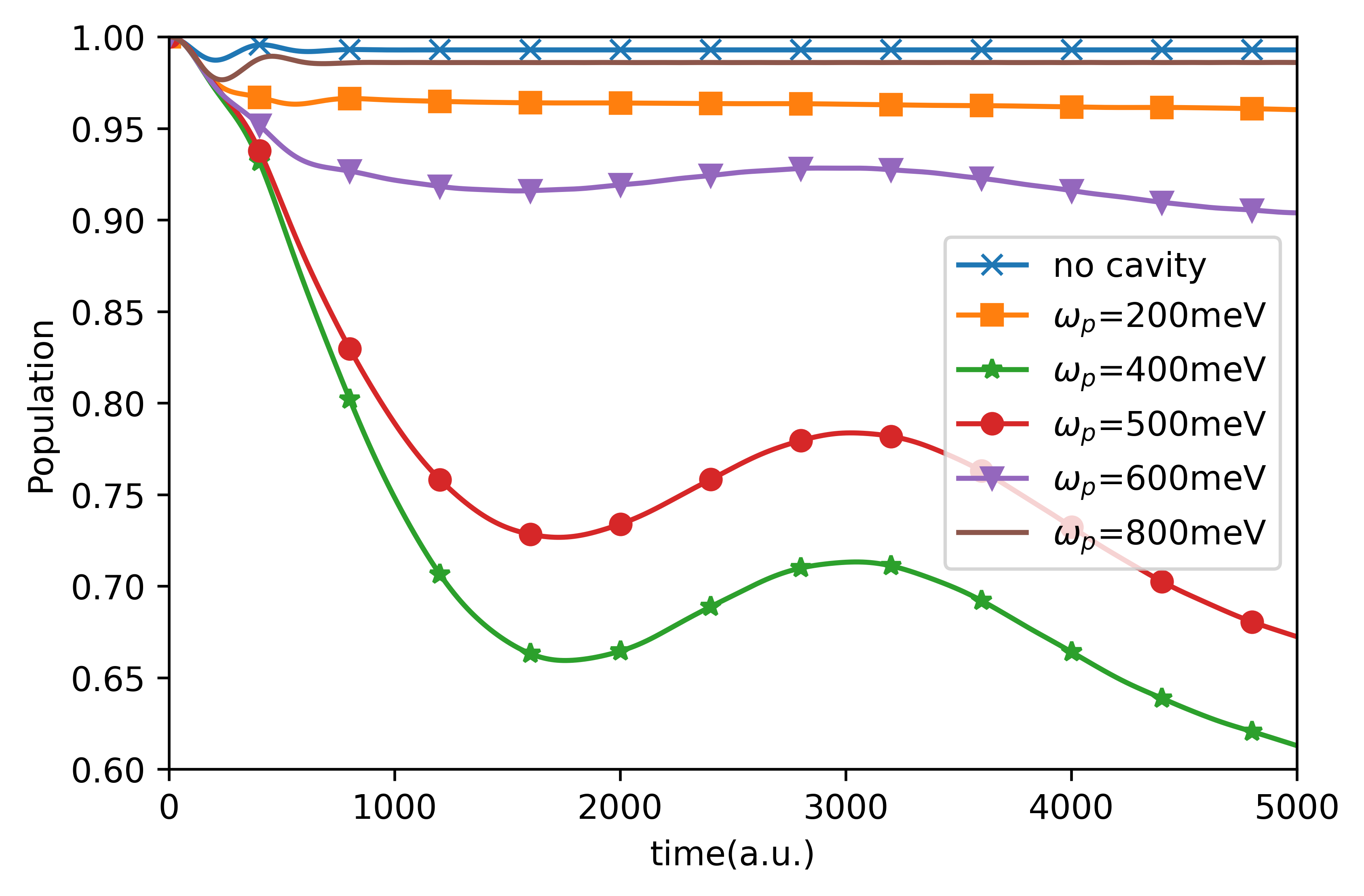}
\caption{Time-dependent population of the ($\pi\pi^*$) state during the early time relaxation after photoexcitation of the carotenoid-porphyrin-C60 molecular triad solvated in tetrahydro-
furan (THF) prepared in a geometry relaxed bent configuration in an optical cavity. The cavity frequency is changed in the $\omega_p=$~200--400~meV range as indicated in the inset. The light-matter interaction strength is set at $g_p=$20~meV. The temperature is set at $T=300~K$. The calculations are based TT-TFD simulations with a propagation time-step $\tau=1~a.u.$}
\label{fig:triad_cav_scan}
\end{figure}

Figure~\ref{fig:triad_cav_scan} shows the effect of changing the cavity resonance frequency on the intramolecular electron transfer dynamics. While light of all frequencies enhance the transfer, the ultrafast transfer efficiency reaches its peak when $\hbar\omega_p$ is around 400~meV. It is worth noting that this optimal cavity efficiency matches relatively well with that predicted by the FGR rate theory\cite{Saller2023} (around 510~meV), despite the fact that the ultrafast dynamics does not exactly follow the exponential relaxation predicted by the FGR rate theory. 

\begin{figure}[H]
\includegraphics[scale=0.8]{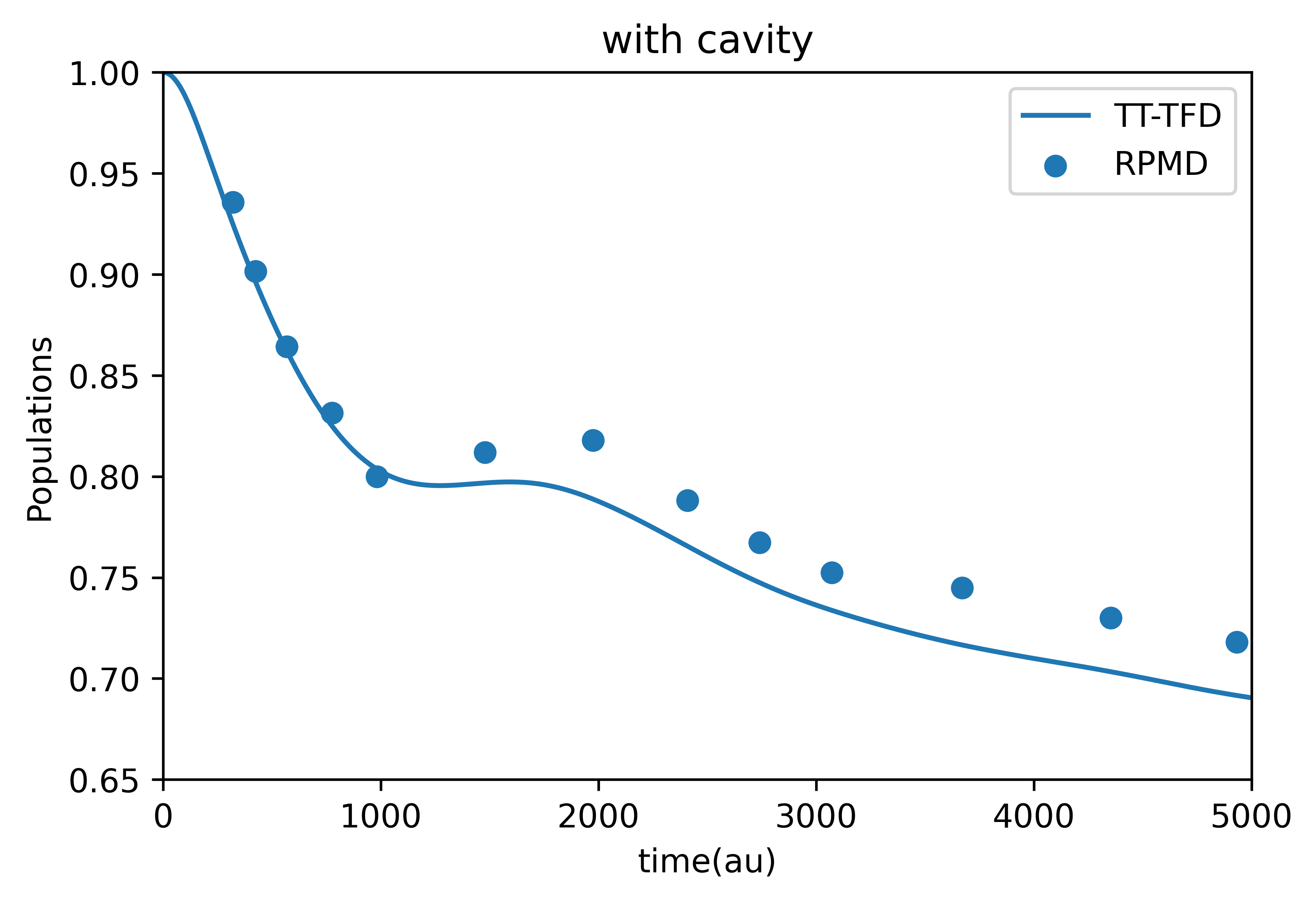}
\includegraphics[scale=0.8]{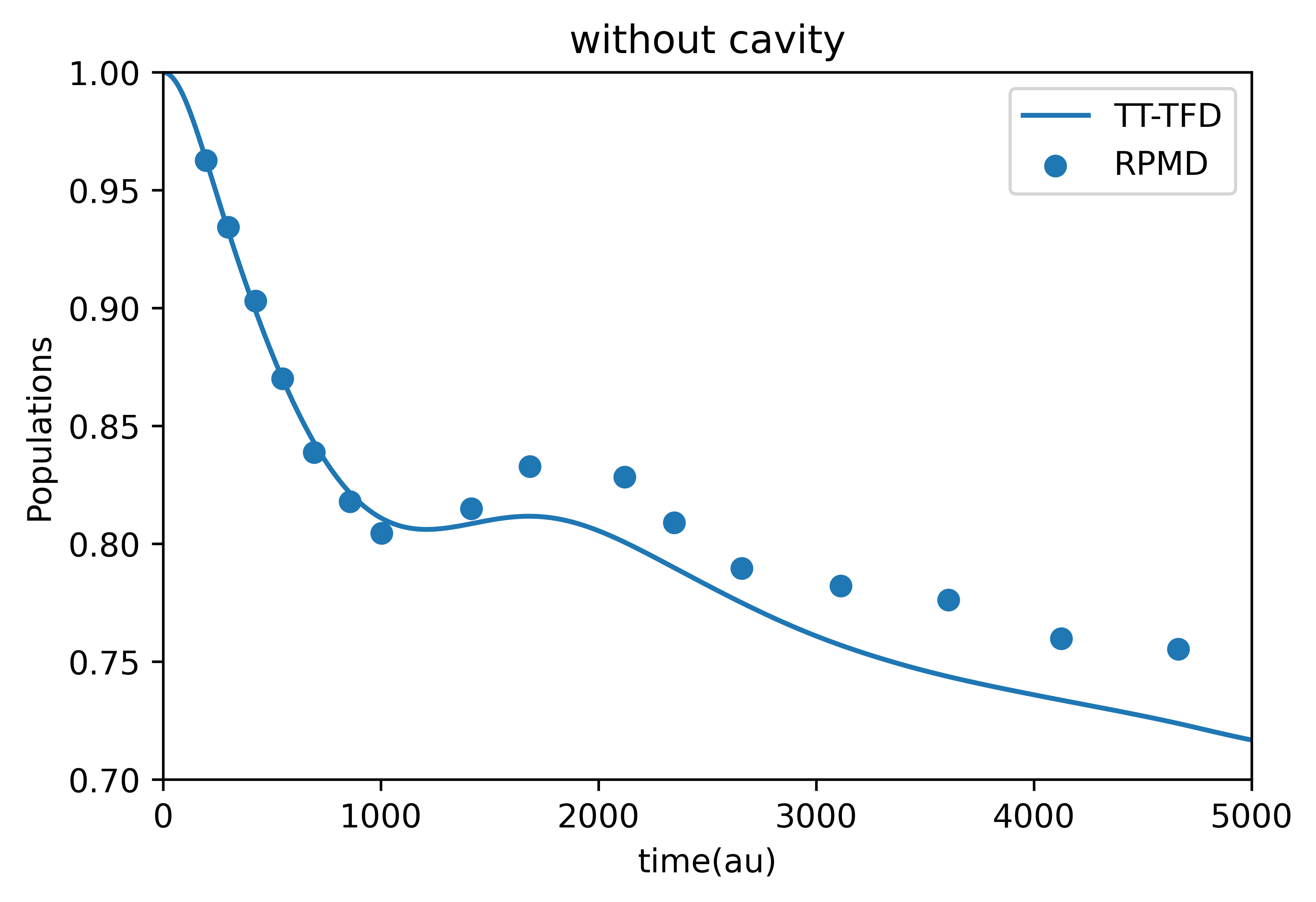}
\caption{Comparison between population dynamics generated by TT-TFD and by RPMD for Model III in Ref.~\citenum{chowdhury21}. (a): donor population with a cavity, (b): donor population without a cavity.}
\label{fig:TFD/RPMD}
\end{figure}

Figure~\ref{fig:TFD/RPMD} shows the comparison of the population dynamics described by numerically exact TT-TFD calculations and approximate results based on Ring-Polymer Molecular Dynamics (RPMD) simulations, reported in Ref.~\citenum{chowdhury21}. Figure~\ref{fig:TFD/RPMD} shows that RPMD agrees qualitatively with numerically exact calculations although exhibits noticeable quantitative deviations. 

\begin{figure}[H]
\includegraphics[scale=0.8]{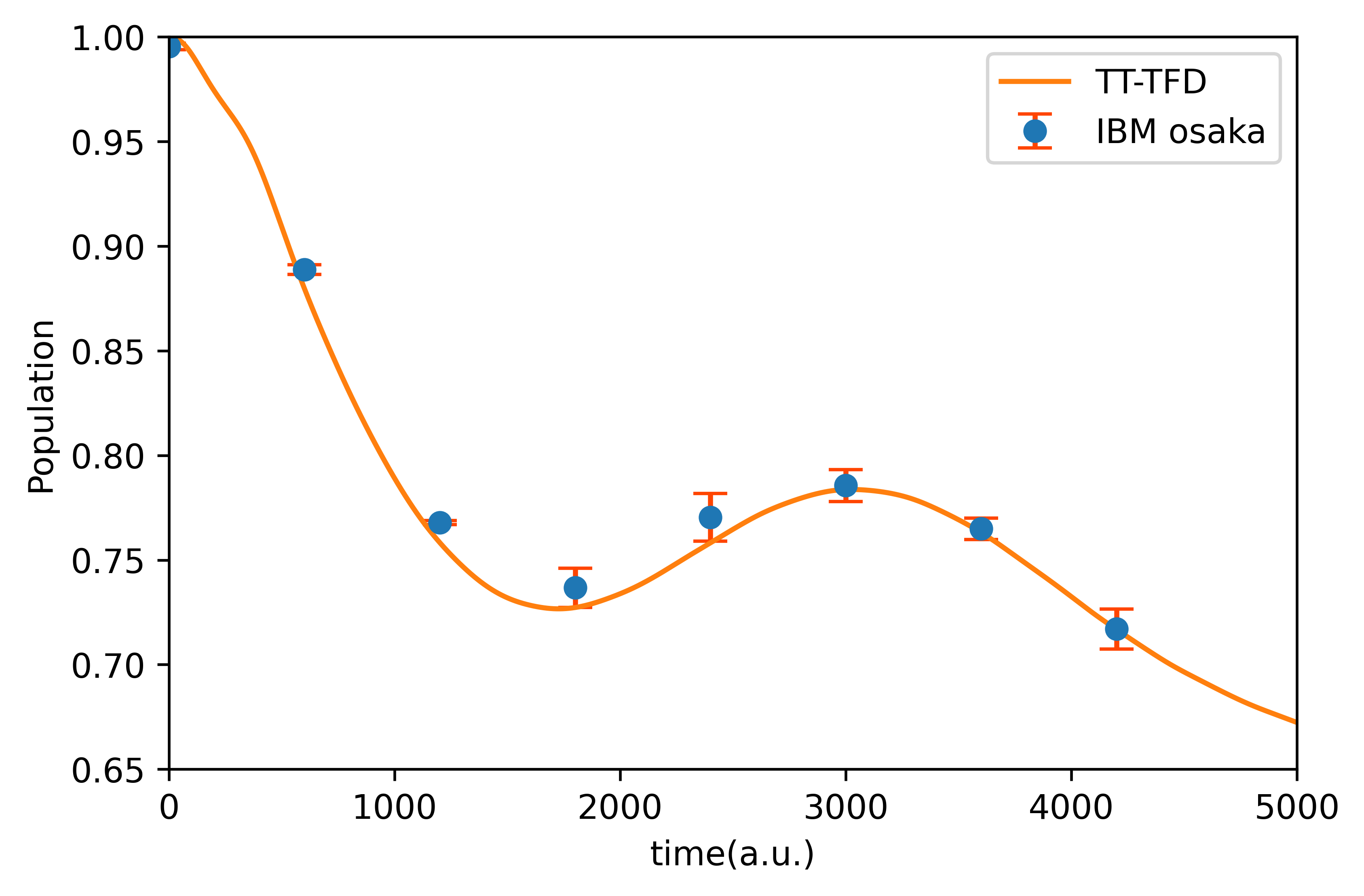}
\caption{Time-dependent population of the ($\pi\pi^*$) state during the early time relaxation after photoexcitation of the carotenoid-porphyrin-C60 molecular triad solvated in tetrahydro-
furan (THF) prepared in a geometry relaxed bent configuration in an optical cavity obtained with the IBM Osaka quantum computer. The cavity frequency is fixed at $\omega_p=$~500~meV. The light-matter interaction strength is set at $g_p=$20~meV. Matrix-free measurement mitigation is performed for readout error mitigation, as implemented in Qiskit. The blue dots are averages over 10 calculations following Eq.~\eqref{diamatvec}, each obtained with 2000 measurements. The error bar shows the highest and lowest values among the 10 runs.
}
\label{fig:triad_cav_qc}
\end{figure}

Figure~\ref{fig:triad_cav_qc} shows the results of calculations of intramolecular electron transfer dynamics performed with the IBM Osaka quantum computer. The results show that the quantum computing scheme is able to produce accurate results within about 1 percent error. 

\section{Conclusions}
We have studied the optical cavity-modulated intermolecular electron transfer dynamics using a rigorous methodology that combines the numerically exact TT-TFD method with open quantum simulations based on dilation. We have shown that the method offers a practical and accurate pathway for quantum computing simulations of cavity modulated quantum reaction dynamics in molecular systems. The simulations are based on a single model molecular triad in the infinitely dilute limit, and therefore neglect collective effects that become important at higher concentration. Nevertheless, our simulation results have shown that the cavity modulation can be significant even in the infinitely dilute limit, strongly modulating the charge transfer dynamics in the ultrafast timescale. 

\section{Acknowledgments}
The authors acknowledge support from the NSF grant 2124511 [CCI Phase I: NSF Center for Quantum Dynamics on Modular Quantum Devices (CQD-MQD)].


\bibliography{main}
\end{document}